\documentclass[12pt,preprint]{aastex}

\begin{document}

\title{The Jet Power and Emission Line Correlations of Radio Loud Optically Selected Quasars}
\author{Brian Punsly\altaffilmark{1}} \and\author{Shaohua Zhang\altaffilmark{2}}
\altaffiltext{1}{4014 Emerald Street No.116, Torrance CA, USA 90503
and ICRANet, Piazza della Repubblica 10 Pescara 65100, Italy,
brian.punsly@verizon.net or brian.punsly@comdev-usa.com}
\altaffiltext{2}{CAS Key Laboratory for Research in Galaxies and Cosmology,
Department of Astronomy, University of Sciences and Technology of China,
Hefei, Anhui, 230026, China}

\begin{abstract}
In this Letter, the properties of the extended radio emission form
SDSS DR7 quasars with $0.4<z<0.8$ is studied. This low redshift
sample is useful since any corresponding FIRST radio observations
are sensitive enough to detect extended flux in even the weakest FR
II radio sources. In the sample, 2.7\% of the sources have
detectable extended emission on larger than galactic scales ($>$ 20
- 30 kpc). The frequency of quasars with FR II level extended radio
emission is $\approx 2.3\%$ and $>0.4\%$ of quasars have FR I level
extended radio emission. The lower limit simply reflects the flux
density limit of the survey. The distribution of the long term time
averaged jet powers of these quasars, $\overline{Q}$, has a broad
peak $\sim 3\times 10^{44}$ ergs/sec that turns over below below
$10^{44}$ ergs/sec and sources above $10^{45}$ ergs/sec are
extremely rare. It is found that the correlation between the
bolometric (total thermal) luminosity of the accretion flow,
$L_{bol}$, and $\overline{Q}$ is not strong. The correlation of
$\overline{Q}$ with narrow line luminosity is stronger than the
correlation with broad line luminosity and the continuum luminosity.
It is therefore concluded that previous interpretations of
correlations of $\overline{Q}$ with narrow line strengths in radio
galaxies as a direct correlation of jet power and accretion power
have been overstated. It is explained why this interpretation
mistakenly overlooks the sizeable fraction of sources with weak
accretion luminosity and powerful jets discovered by Ogle et al
(2006).
\end{abstract}

\section{Introduction}
Three major questions regarding relativistic jets in radio loud
extragalactic radio sources that are of fundamental interest are,
"how powerful is the typical radio jet?", "how frequently do quasars
produce long term jets?" and "how is the jet power related to the
accretion flow thermal luminosity?". The signature of long term
powerful jets are the radio lobes of Fanaroff-Riley II (FR II) radio
sources. Quasars with double radio lobes are conspicuous, but quite
rare. Thus, radio selected samples are typically used to address the
questions above since sufficient sample size is required for
statistical merit. Most radio loud sources \footnote{The radio
loudness, $R$, is usually defined as a 5 GHz flux density 10 times
larger than the $4400 \AA$ flux density, $R=S_{5
\mathrm{GHz}}/S_{4400 \AA}<10$ \citep{kel69}.} do not display radio
lobes with the resolution and dynamic range of FIRST \citep{dev06}.
Thus, in order to find the frequency of the FR II phenomenon in
quasars and the complete distribution of their kinetic luminosity,
one must have the ability to detect lobe emission that is below the
FR I/FR II divide. The FR I/FR II distinction is both a
morphological and luminosity classification. In terms of long term
average jet power, $\overline{Q}$, this approximate boundary is at
$\overline{Q} \approx 5 \times 10^{43}$ ergs/sec
\citep{raw91,pun08}. This divide corresponds to a 1.4 GHz flux
density $< 12.5$ mJy at $z < 0.8$, well within the sensitivity of
the FIRST radio survey, even if the emissivity is distributed over
10 beamwidths (see equation (1), below). Thus, the extant databases
are now at a point where the questions above can be addressed
rigorously.
\par We intend to study $\overline{Q}$
with a carefully selected, large sample of SDSS DR7 quasars
($0.4<z<0.8$) that is cross-referenced to the FIRST
database. Selecting the sources in this way reveals prominent narrow
lines ([OIII] $\lambda$5007 and [OII] $\lambda$3727) and two broad
lines H$\beta$ $\lambda$4861 and Mg II $\lambda$2798 in the SDSS spectral
window for all objects. This allows us to compare the various derived
properties in the context of a single sample of objects in a way
that removes the uncertainties associated with sample selection
biases if a different set of objects and lines are used for the
analysis depending on redshift. The cross-correlations between the
different estimators of the photo-ionizing continuum allow us to
segregate out the effects of Doppler beaming and line emission
created by jet induced shocks, thereby reducing systematic
uncertainties in the bolometric luminosity, $L_{bol}$, the broadband
thermal luminosity from IR to X-ray, of the active galactic nucleus
(AGN).
\par In the next section, the sample is described as well as the extended
radio flux data reduction and the optical spectra data reduction.
The third section summarizes the distribution of $\overline{Q}$ that is found.
The fourth section discusses the correlations of $\overline{Q}$ with the
optical/UV spectral parameters.

\begin{figure}
\includegraphics[scale=0.75]{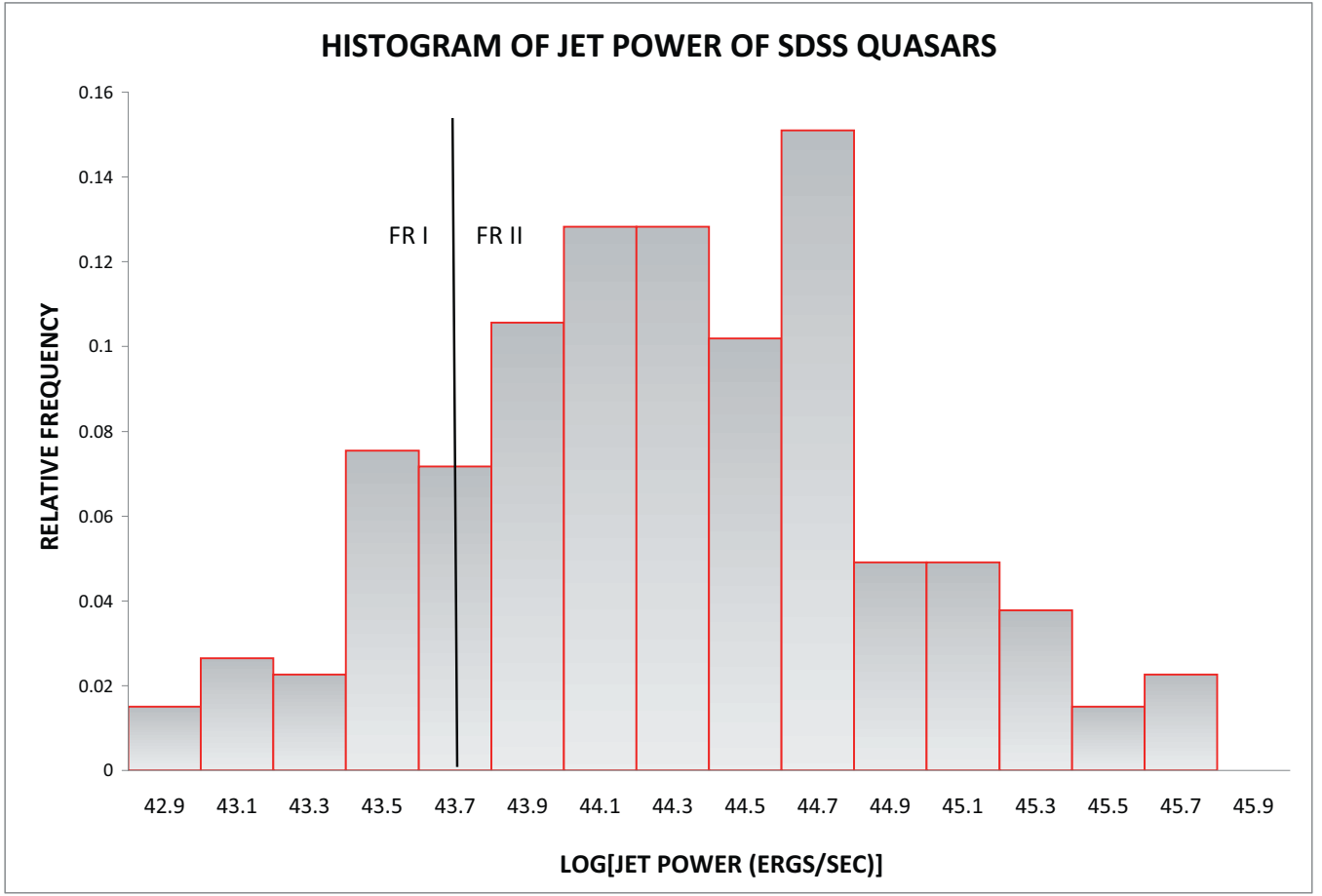}
\caption{The histogram of the quasar long term time averaged jet
power, $\overline{Q}$, computed by means of equations (1) and (2).}
\end{figure}
\section{Sample Description and Data Extraction Methods}
In order to determine the dependence of line luminosity on the
thermal continuum in a quasar, we constructed a sample of SDSS DR7
quasars with $0.4<z<0.8$ and spectra with a median S/N $>$ 7, this
yielded 10069 AGNs. This base sample has been analyzed previously
for other purposes in \citet{pun10,pun11}. The SDSS DR7 data was
cross-referenced to the FIRST database. Of these, there were 9270
sources found within these FIRST fields that comprise our sample.
Most of the radio loud sources did not have extended radio flux with
the dynamic range and resolution of FIRST. Indicating a sub-galactic
compact structure or very weak, FR I level, extended emission. The optical spectra
were reduced as in \citet{pun11} yielding the following for every quasar:
the optical/UV continuum luminosity from 5100 $\AA$ to 3000 $\AA$, $L_{\mathrm{cont}}$,
the luminosity of the broad components of H$\beta$ $\lambda$4861, $L_{H\beta}$, and Mg II $\lambda$2798, $L_{Mg II}$,
and the luminosity of the narrow lines [OIII] $\lambda$5007, $L_{OIII}$, and [OII] $\lambda$3727, $L_{OII}$.
\par The spatial resolution associated with the
interferometer beamwidth of the FIRST survey
is $\approx$ 20 kpc at z = 0.4 and $\approx$ 30 kpc at z = 0.8. The optically thin extended flux density
on scales of this size and larger can be used to estimate the long term time averaged kinetic energy
flux in a jet, $\overline{Q}$ \citep{wil99}. FIRST observations are at a fairly low frequency, 1.4 GHz, so
they are sensitive to steep spectrum optically thin emission. Resolved emission at 1.4 GHz
is very likely to be optically thin \citep{ant85,mur93}. Thusly motivated, we want to extract this extended flux from
the FIRST radio database for the SDSS quasars.
\par The first step was to cross-reference all FIRST sources within 30" of the SDSS sources.
Each one of these radio fields was inspected visually. Diffuse
emission that is roughly concentric with either the optical quasar
position, or an unresolved radio core located within 3" of the
optical position, is considered extended emission of the quasar.
This would be core - halo type emission. Components within 10" of
the optical/radio core are considered extended emission. Distant
components located 10" to 70" from the quasar/radio core position
are considered to be associated with the quasar if there is an
extension or jet pointing towards the direction of the optical
quasar/radio core position, or if there is a second distant
component located on a line that passes close to this optical
quasar/core position (i.e., a classical triple). We also considered
partially resolved sources in the image plane, quantifying the
extended emission as the integrated flux minus the peak flux. Other
configurations were not considered extended flux associated with the
quasar (either chance background sources or core flux).
\par The most accurate estimates of $Q$ should use an isotropic
estimator such as the radio lobe flux. The sophisticated calculation
of the jet kinetic luminosity in \citet{wil99} incorporates
deviations from the overly simplified minimum energy estimates into
a multiplicative factor $f$ that represents the small departures
from minimum energy, geometric effects, filling factors, protonic
contributions and low frequency cutoff. The quantity, $f$, is argued
to be constrained between 1 and 20. In \citet{blu00}, it was further
determined that $f$ is most likely in the range of 10 to 20. Thus,
choosing a value of $f=15$, \citet{pun05} converted the analysis of
\citet{wil99} to the formula in equations (1) and (2), even though it is just a time
average. It is assumed throughout this Letter that $H_{0}$=70 km/s/Mpc, $\Omega_{\Lambda}=0.7$ and $\Omega_{m}=0.3$.
This formula is an isotropic method that allows one to
convert 151 MHz flux densities, $F_{151}$ (measured in Jy), into
estimates of $\overline{Q}$ (measured in ergs/s):
\begin{eqnarray}
&& \overline{Q} \approx 1.1\times
10^{45}\left[(1+z)^{1+\alpha}Z^{2}F_{151}\right]^{\frac{6}{7}}\mathrm{ergs/s}\;,\\
&& Z \equiv 3.31-(3.65) \nonumber \\
&&\times\left(\left[(1+z)^{4}-0.203(1+z)^{3}+0.749(1+z)^{2}
+0.444(1+z)+0.205\right]^{-0.125}\right)\;,
\end{eqnarray}
where $F_{151}$ is the total optically thin flux density from the
lobes (i.e., \textbf{no contribution from a Doppler boosted jet or
the radio core}). The spectral index is defined by the convention,
$F_{\nu} \sim \nu^{-\alpha}$. The appropriate application of this
equation requires that one must extricate the diffuse lobe emission
from the Doppler boosted core and jet. The expression, (1), requires
151 MHz flux densities, so we extrapolate the 1.4 GHz extended flux
from the FIRST observation. Without further knowledge, for lobe
emission $\alpha \approx 0.8$ is the most typical value found in
\citet{kel70}, and this value is used to extrapolate the 1.4 GHz
flux density to 151 MHz. These extrapolations were also
cross-referenced to any available low frequency data in NED for a
consistency check. There is no evidence that a significant amount of
flux was resolved out by the FIRST beam size. For a discussion of
the errors implicit in equations (1) and (2), see the discussions of
\citet{fer10,blu00,wil99}. The independent derivation in
\citet{pun05} indicates that most estimates in a large sample will
be accurate to within a factor of 2.
\section{Long Term Time Averaged Jet Power} Figure 1 is a histogram
of $\overline{Q}$ for the 266 quasars with detected extended radio
emission. Notice that there is a partition associated with an
approximate FR I/FR II dividing line at $\overline{Q}= 5\times
10^{43}\mathrm{ergs/sec}$ \citep{raw91,pun08}. The histogram shows
that the FIRST sensitivity in this redshift range is capable of
detecting FR I level extended emission. There are 220 FR II sources
above the FR I/FR II divide and 46 sources below the divide. This
corresponds to 2.3\% of the optically selected quasars have FR II
level extended emission and $>0.4\%$ have FR I level extended
emission. The fraction that are FR I sources would certainly
increase with deeper radio observations. The fraction that are FR II
would also likely increase with deeper radio imaging (change some FR
I levels to FR II levels) as faint diffuse flux is likely swamped by
the FIRST rms noise. The frequency of 2.3\%  for the FR II quasars
is consistent with the value of 1.7\% deduced by \citet{dev06} since
their sample has a mean of $z \approx 1.3$, so many faint, diffuse
FR II sources (that are prominent in Figure 1) would be difficult or
impossible to detect with FIRST sensitivity at these high redshifts,
hence their lower deduced rate of occurrence. The significant
fraction of FR I extended flux densities is consistent with the
discovery of quasars with FR I extended radio emission in
\citet{gow84,blu01}.
\par A major advantage of this study is that the peak of the quasars extended
luminosity distribution is resolved in Figure 1 due to the FIRST
sensitivity in this redshift range. It shows that the most frequent
luminosity (for sources with extended emission) is definitely at the
FR II level. The peak level of the histogram is $\gtrsim
10^{44}\mathrm{ergs/sec}$ with marginal evidence of a second peak at
$\approx 5 \times 10^{44}\mathrm{ergs/sec}$. Only 33 sources have
$\overline{Q} > 10^{45}\mathrm{ergs/sec}$,
 $\approx$ 0.3\%. Although powerful FR II quasars ($\overline{Q} > 10^{45}\mathrm{ergs/sec}$)
are prominent in low frequency radio surveys such as the 7C survey,
\citet{wil99}, these sources are extremely rare, perhaps the rarest
subclass of quasar - an order of magnitude rarer than low
ionization broad absorption line quasars \citep{zha10}. Whatever
physical phenomenon creates these large values of $\overline{Q}$
must be very difficult to achieve and maintain for long periods of time in a quasar
environment.

\section{Correlations Between Accretion Luminosity and Jet Power} The blue continuum luminosity is the most
basic signature of the thermal emission from the quasar, so it is
the most commonly used quantity for estimating $L_{bol}$. Perhaps
the most popular bolometric correction is the simple one proposed by
\citet{kas00}, $L_{bol}\approx 9\lambda L_{\lambda}(5100\,\AA)$.
Clearly, using a portion of the optical/UV continuum is more
accurate than a single point and we have that at our disposal,
$L_{\mathrm{cont}}$. In \citet{pun11}, in another study of these
10,069 SDSS sources, the \citet{kas00} relation was converted to
\begin{equation}
L_{bol} \approx 15 L_{\mathrm{cont}}\;.
\end{equation}
$L_{\mathrm{cont}}$ is the most direct way to estimate $L_{bol}$.
One can also estimate $L_{bol}$ from the spectral luminosity at 3000
$\AA$ for the sources in this sample. The average spectral index of
the optical/UV continuum for the radio quiet sources in the sample
of 10,069 quasars was found in \citet{pun11} to be $\alpha \approx 0.55$,
where $L_{\nu} \sim \nu^{-\alpha}$. Thus, the \citet{kas00}
estimator can also be realized as
\begin{equation}
L_{bol} \approx 7.1 \lambda L_{\lambda}(3000\, \AA) \;.
\end{equation}

It is interesting to see if there is a strong relationship between
$L_{bol}$ and $\overline{Q}$ as has been inferred for radio galaxies
in \citet{raw91,wil99}. An uncertainty in equation (3), in the the
context of radio sources, is the high frequency tail of the
synchrotron emitting jet and its potential to dilute the pure
$L_{\mathrm{cont}}$ from the accretion flow. Our carefully selected
sample ameliorates these concerns. We have other reliable surrogates
for $L_{bol}$. The synchrotron tail dies of rapidly with frequency
in the optical/UV and the big blue bump from the accretion disk
becomes more prominent in the UV \citep{mal86}. The spectral index
of $L_{\nu}$ of the optical/UV synchrotron tail in quasars is
typically, $\alpha \sim 2$, compared to the accretion spectral
luminosity with $\alpha \approx 0.55$ for the 10,069 sources in the
SDSS quasar sample \citep{smi88,pun11}. Thus, the ratio of the
synchrotron tail luminosity ($\lambda L_{\lambda}$) to the accretion
luminosity is typically $\gtrsim 2$ times larger at 5100 $\AA$ than
at 3000 $\AA$. Consequently, if synchrotron dilution is significant,
there should be a noticeable difference in the correlations of
quasar parameters with $L_{\mathrm{cont}}$ and those with $L(3000
\AA)$. Furthermore, it was shown in \citet{pun11} that the broad
components of Mg II and in particular H$\beta$ are excellent
surrogates for the underlying accretion disk luminosity. These
spectral parameters are available for all sources in our sample.
Figure 2 shows that $L_{H\beta}$ is an excellent surrogate for
$L_{\mathrm{cont}}$ for our radio sources as well. The best fit
relationship displayed at the top of the plot is nearly linear. The
tight correlation is evidence that Doppler beaming of the high
frequency tail of the synchrotron jet is not a major contaminant to
the continuum luminosity measured by SDSS in a statistical sense.

\par We explore the correlations between the spectral features and $\overline{Q}$ in
the correlation matrix, Table 1.
\begin{table}
\caption{Spearman Rank Correlation Matrix for the Quasars with Extended Emission}
\begin{tabular}{cccccccc} \tableline \rule{0mm}{3mm}
                           &   $L_{\mathrm{cont}}$   & $L(3000 \AA)$  & $L_{H\beta}$ & $L_{Mg II}$  & $L_{O III}$ & $L_{O II}$ & $\overline{Q}$ \\
\tableline \rule{0mm}{3mm}
$L_{\mathrm{cont}}$  &  1 &  0.985  &  0.869 & 0.732 & 0.525 & 0.379  & 0.314   \\
$L(3000 \AA)$            &  0.985 &  1  &  0.871 & 0.775 & 0.522 & 0.367  & 0.313   \\
$L_{H\beta}$                  &  0.869 &  0.871  &  1 & 0.799 & 0.636 & 0.468  & 0.417   \\
$L_{Mg II}$                     &  0.732 &  0.775  &  0.799 & 1 & 0.565 & 0.397  & 0.421   \\
$L_{O III}$                     &  0.525 &  0.522  &  0.636 & 0.565 & 1 & 0.844  & 0.507   \\
$L_{O II}$                      &  0.379 &  0.367  &  0.468 & 0.397 & 0.844 & 1  & 0.467   \\
$\overline{Q}$            &  0.314 &  0.313  &  0.417 & 0.421 & 0.507 & 0.467  & 1   \\

\tableline{\rule{0mm}{3mm}}
\end{tabular}
\end{table}
The statistical metric of choice is the Spearman rank correlation
coefficient in Table 1. As expected, $L_{\mathrm{cont}}$  and
$L(3000 \AA)$ are tightly correlated and their correlations with
$L_{H\beta}$ are also tight (see Figure 2). The most intriguing
aspect of Table 1 and Figure 3 is that the weakest correlation with
$\overline{Q}$ is from the continuum itself and then $L_{H\beta}$.
The correlations of $\overline{Q}$ with the narrow line strengths
are more significant than with the broad lines or the continuum. A
lack of a strong correlation between $\overline{Q}$ and
$L_{\mathrm{cont}}$ is consistent with the right hand panel of
Figure 4 of \citet{fer10} based on a sample of optically selected
radio loud quasars.
\begin{figure}
\includegraphics[scale=0.75]{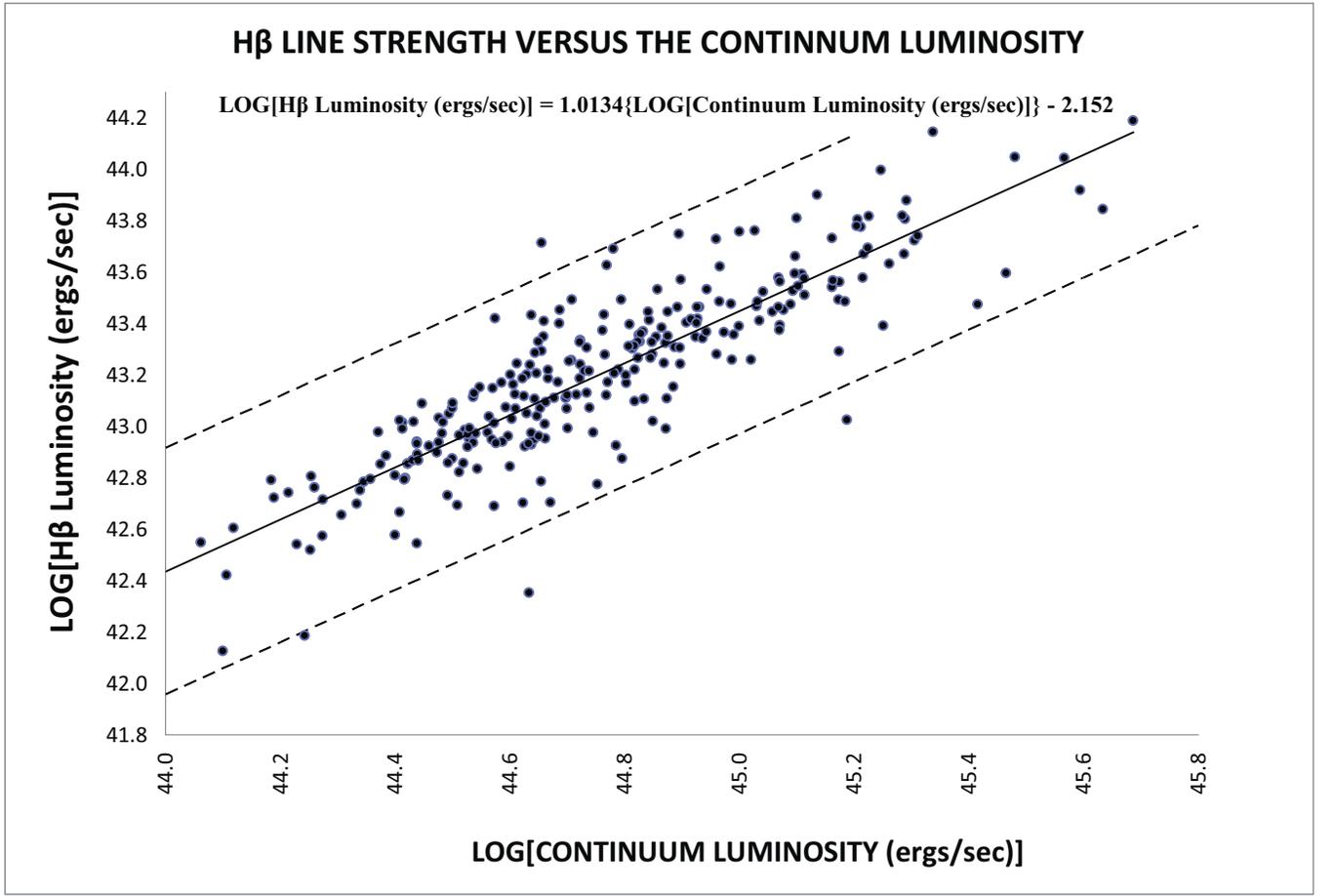}
\caption{The scatter plot of $\log{L_{H\beta}}$ versus
$\log{L_{\mathrm{cont}}}$ for the 266 quasars with extended
luminosity. The dashed lines are the 2.5 $\sigma$ level scatter from the linear
fit that were chosen to highlight the small number of sources below
the trend-line that might be significantly affected by Doppler
boosting.}
\end{figure}

\par Many previous studies have shown a strong correlation between narrow line strengths
and $\overline{Q}$ for radio galaxies
\citep{raw91,wil99,but10,but11,koz11}. However, the more direct
indicators of the continuum in Table 1 (the broad lines and a direct
view of the continuum itself) either did not exist or were hidden by
the dusty torus \citep{ant93}. It was shown in \citet{pun11} that a
narrow line luminosity is not nearly as accurate of a measure of
$L_{bol}$, as $L_{\mathrm{cont}}$ or a broad line luminosity. It
seems that the interpretation of the correlation of $L_{O II}$ with
$\overline{Q}$ for the relatively powerful 7C sources in
\citet{wil99} in terms of a strong correlation between $L_{bol}$ and
$\overline{Q}$ might need some revision. In Table 1, the
correlations of $L_{O II}$ and $L_{O III}$ are stronger with
$\overline{Q}$ than with $L_{\mathrm{cont}}$ or $L(3000 \AA)$. This
suggests that jet propagation is a comparable or more significant source of
excitation energy than photo-ionization for the O II emitting gas in
FR II radio sources. It has been shown that $L_{O III}$ is superior
to $L_{O II}$ as a measure of $L_{bol}$, due to at least two
effects: the dependence on the ionization parameter, \citet{sim98},
and contamination by star formation in host galaxies \citep{hec04}.
Thus, correlations with $L_{O III}$ have a more direct physical
interpretation. For O III, the source of excitation for the narrow
line gas might be roughly an equal mix of photo-ionization energy
and jet induced energy based on the correlation strengths in Table 1.
Thus, the narrow line strength and $\overline{Q}$ are dependent
statistical quantities. So, if a narrow line strength is used as a
surrogate for $L_{bol}$ in a correlation analysis or scatter plot
with $\overline{Q}$ then the probability of large outlier sources
(i.e., large $\overline{Q}$ and small narrow line strength) is
greatly diminished compared to a scatter plot in the $\overline{Q}$
- $L_{bol}$ plane (i.e., large $\overline{Q}$ and small $L_{bol}$)
because $\overline{Q}$ is itself a strong source of narrow line
emitting gas. This interpretation seems to rectify a discrepancy
seen here and elsewhere in the literature. Consider the bimodal
scatter in the IR luminosity/ radio luminosity plane noted by
\citet{ogl06} in their Figure 1. They concluded that significant
$\overline{Q}$ can exist in sources where there is essentially very
little accretion thermal luminosity as well as in sources with large
(quasar-like) $L_{bol}$. Conversely, the correlation between narrow
line strength and $\overline{Q}$ in \citet{raw91,wil99} led to the
conclusion that accretion power was very closely linked to jet
power. Based on Table 1, this difference in interpretation arises
because a strong radio source with a weak ionizing continuum is
still likely to be a major source of narrow line emission, thereby
keeping these sources correlated with narrow line strength - even
though they are not correlated with $L_{bol}$. An important point to
bear in mind is that only a small range of optical luminosity is
probed here, with most sources being within one order of magnitude.
Given the very large scatter in some of these correlations, it is
clear that over such a small range of luminosity, correlation
statistics can give low values. Observations over many orders of
magnitude with the same scatter would give much higher correlation
coefficients, e.g. the correlations of \citet{raw91,wil99}. However,
the scatter in these relationships is still significant, 0.54 dex
\citep{wil00}.
\begin{figure}
\includegraphics[scale=0.35]{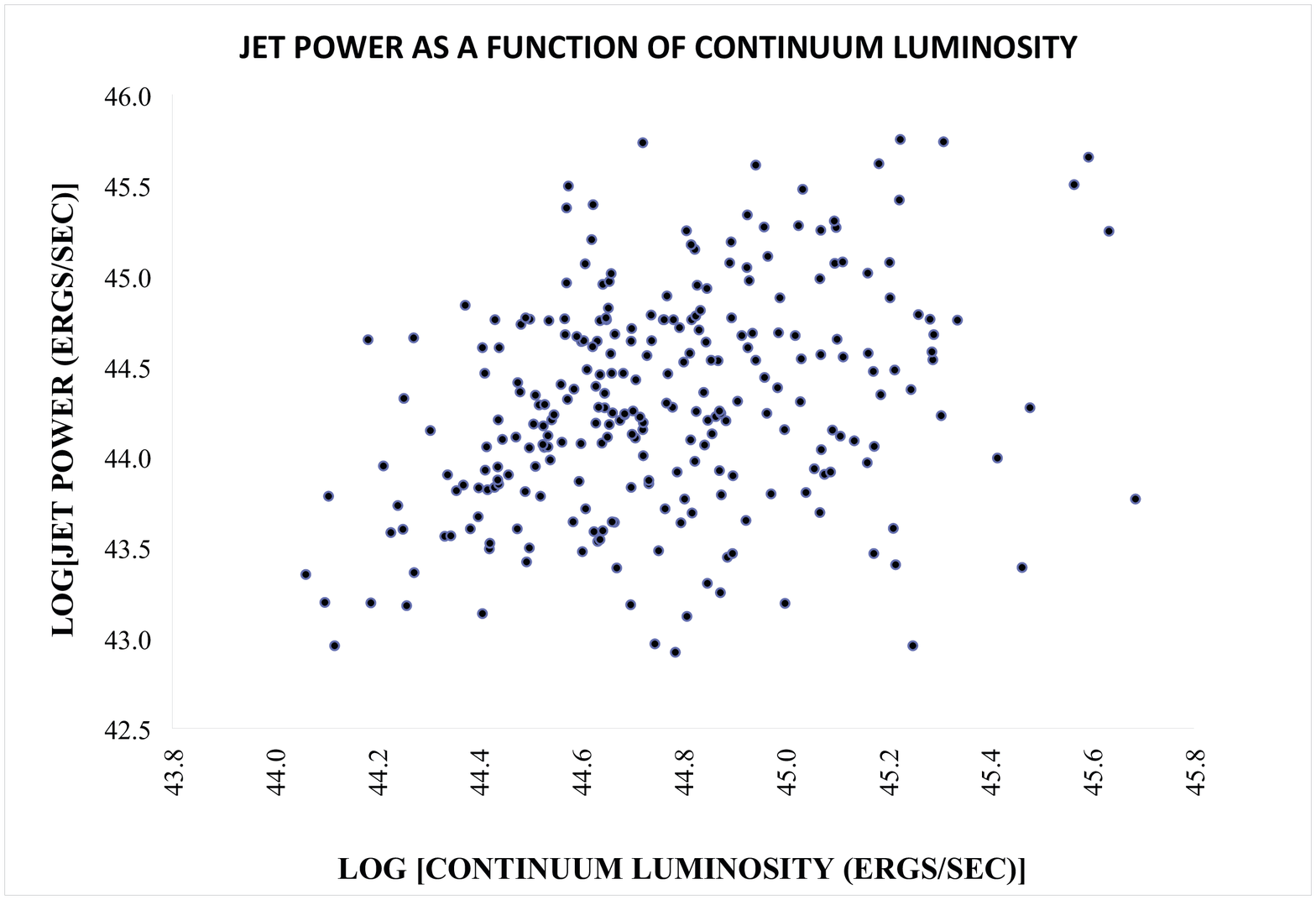}
\includegraphics[scale=0.35]{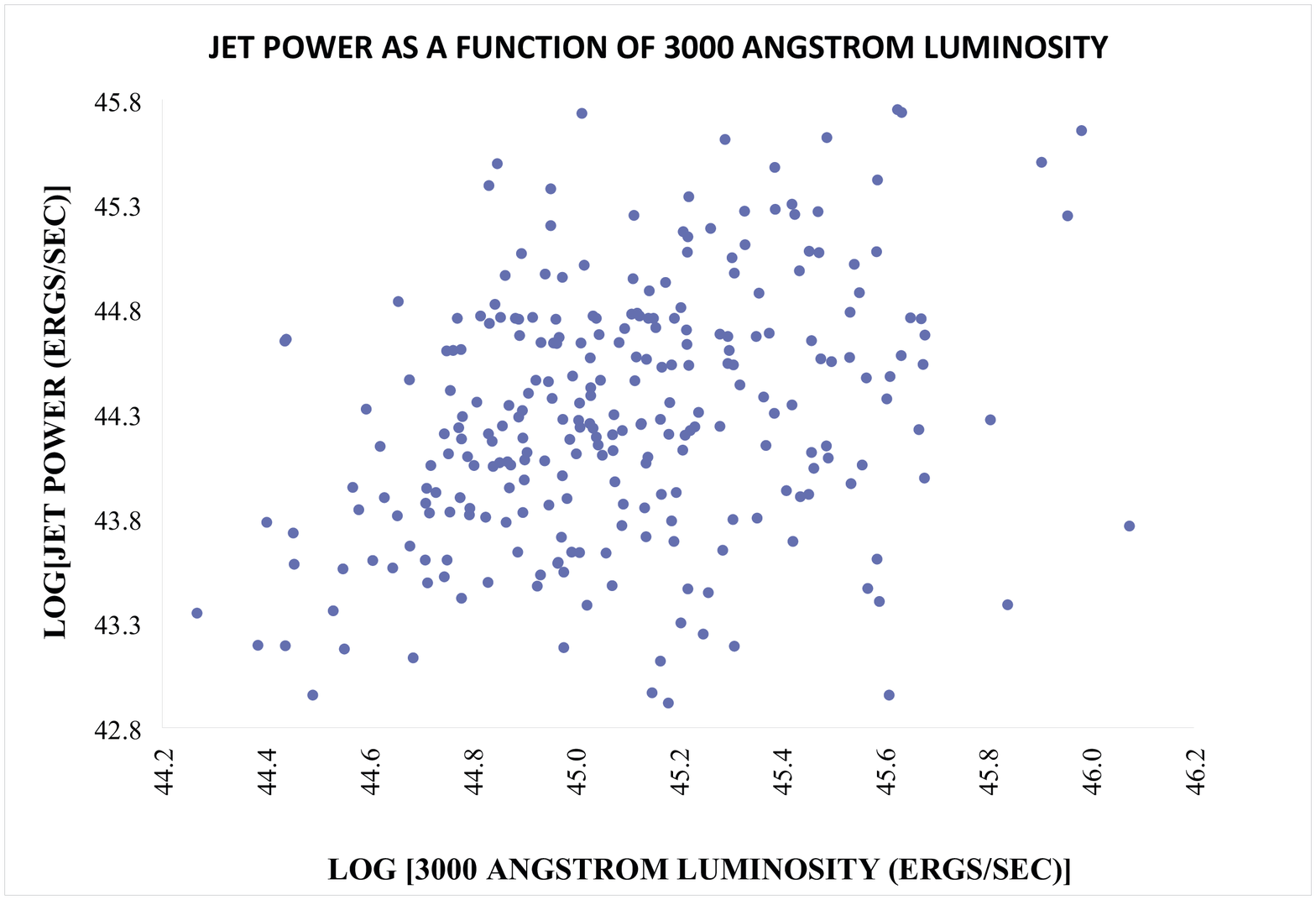}\\
\includegraphics[scale=0.35]{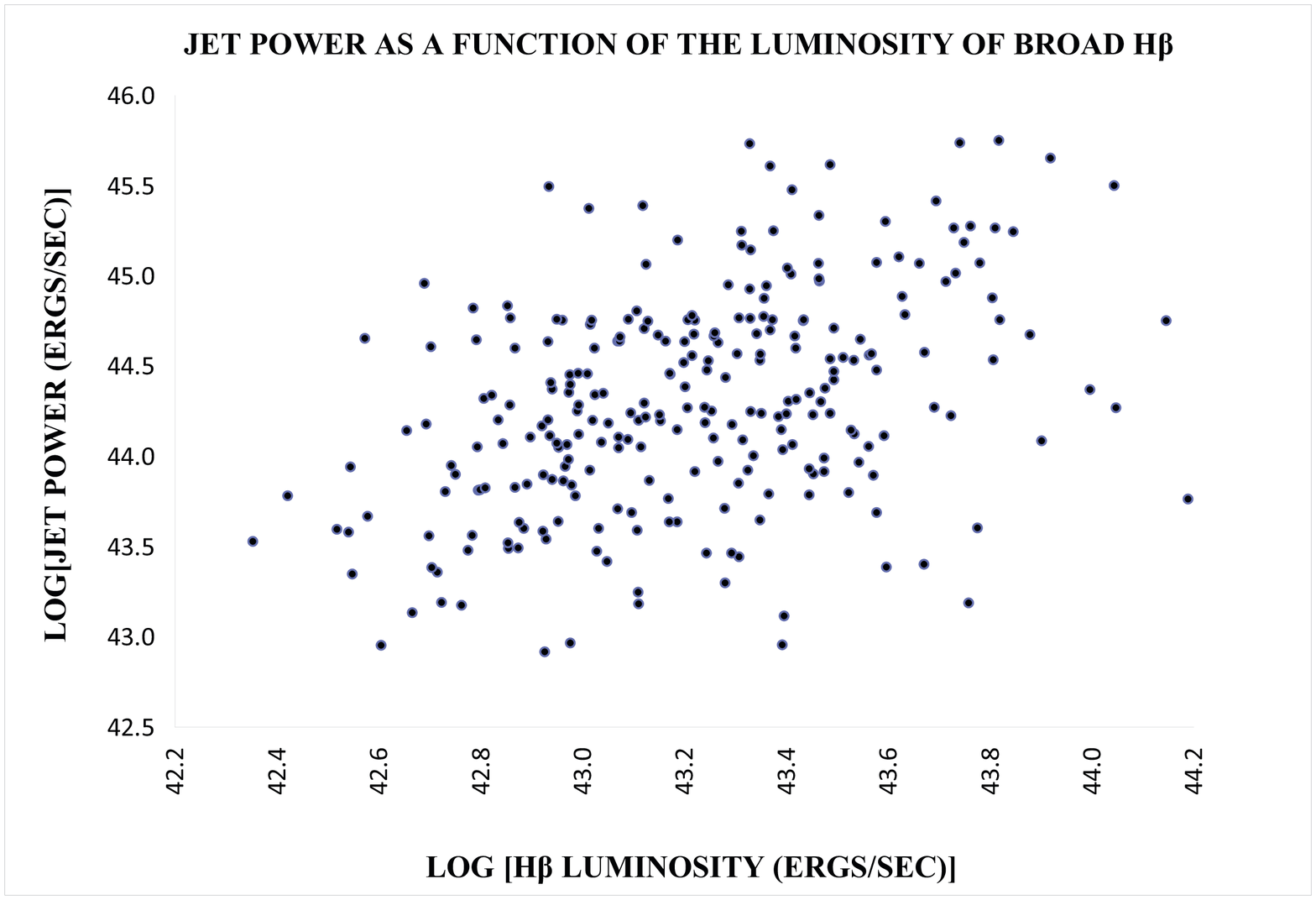}
\includegraphics[scale=0.35]{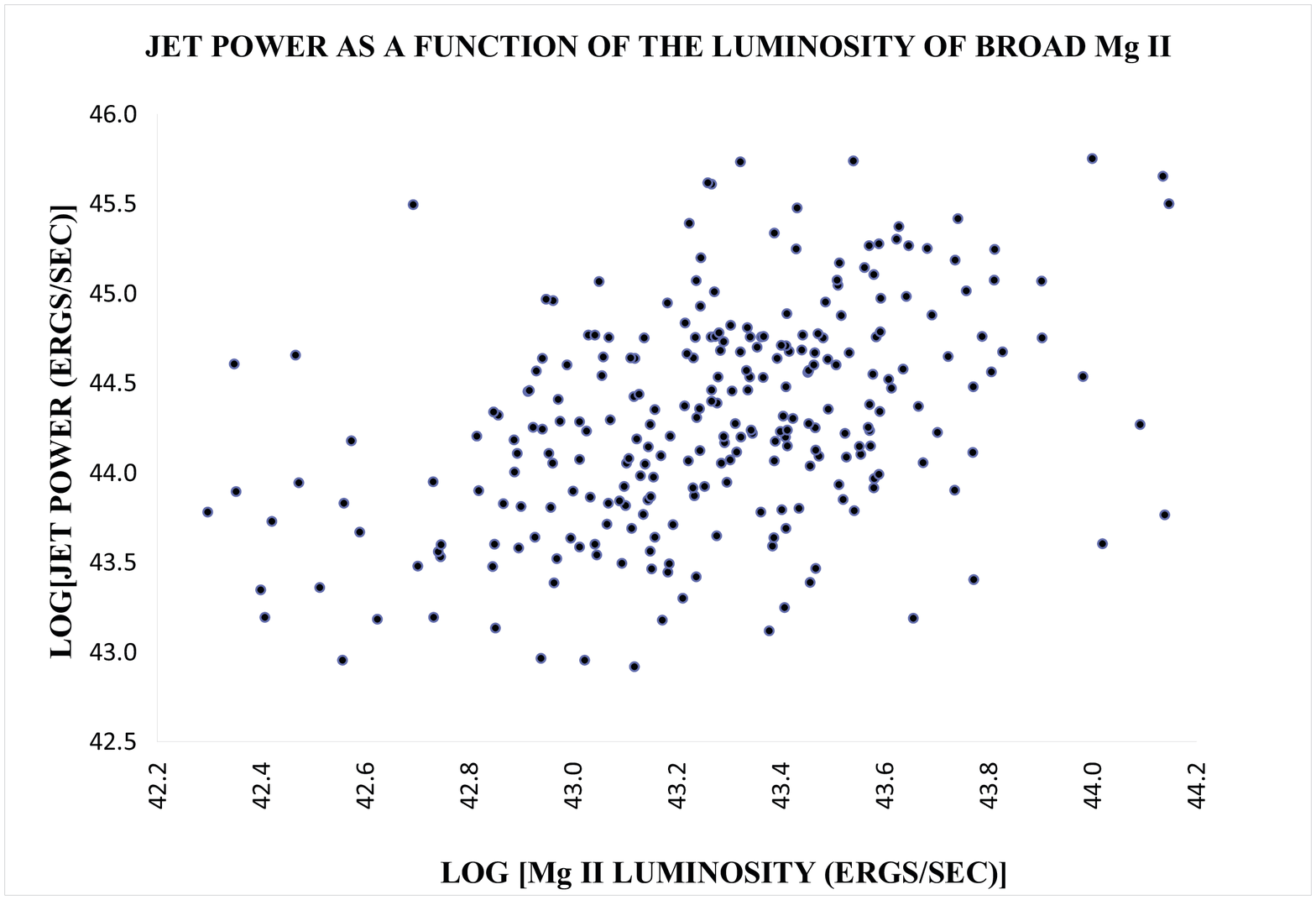}\\
\includegraphics[scale=0.36]{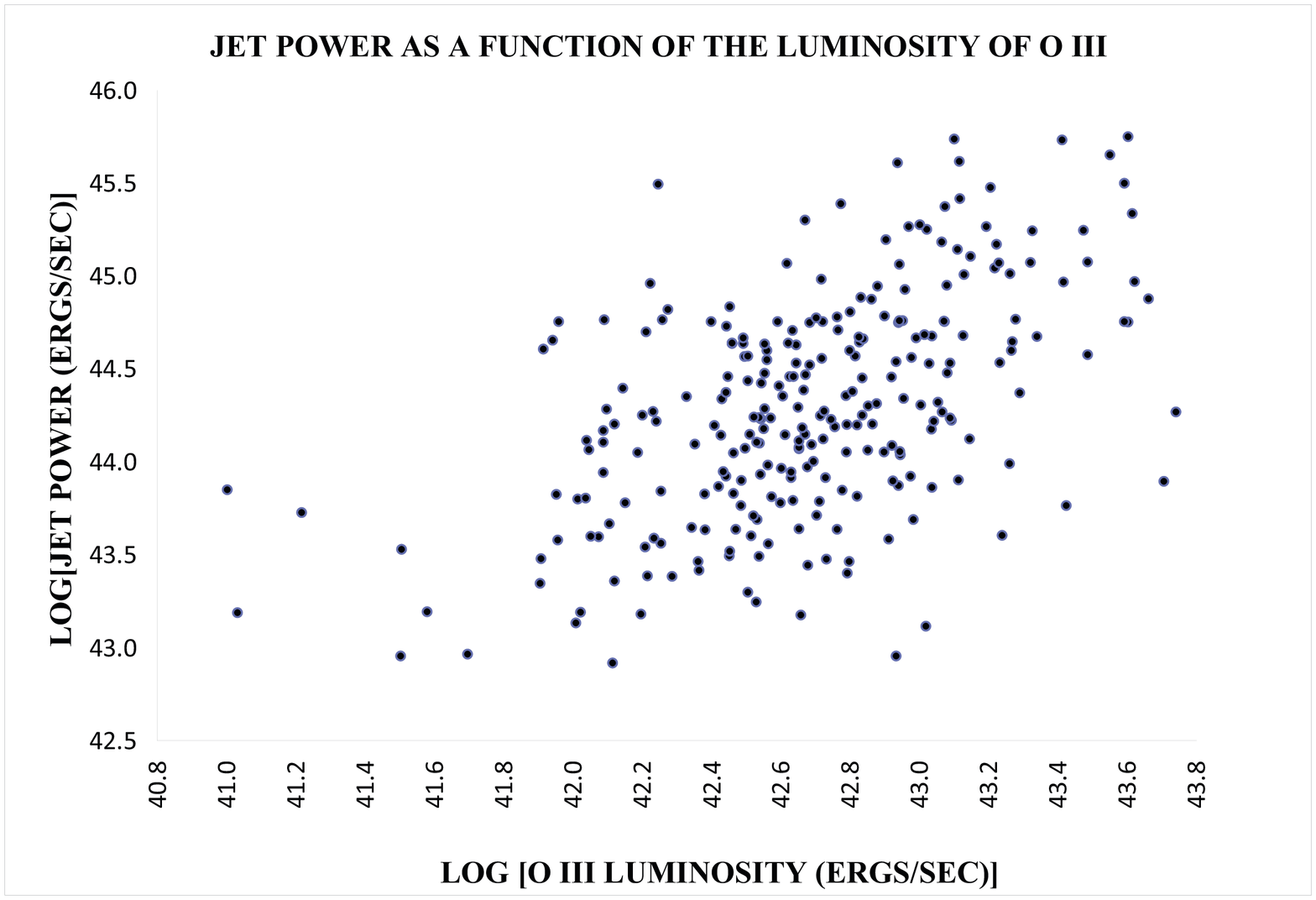}
\includegraphics[scale=0.35]{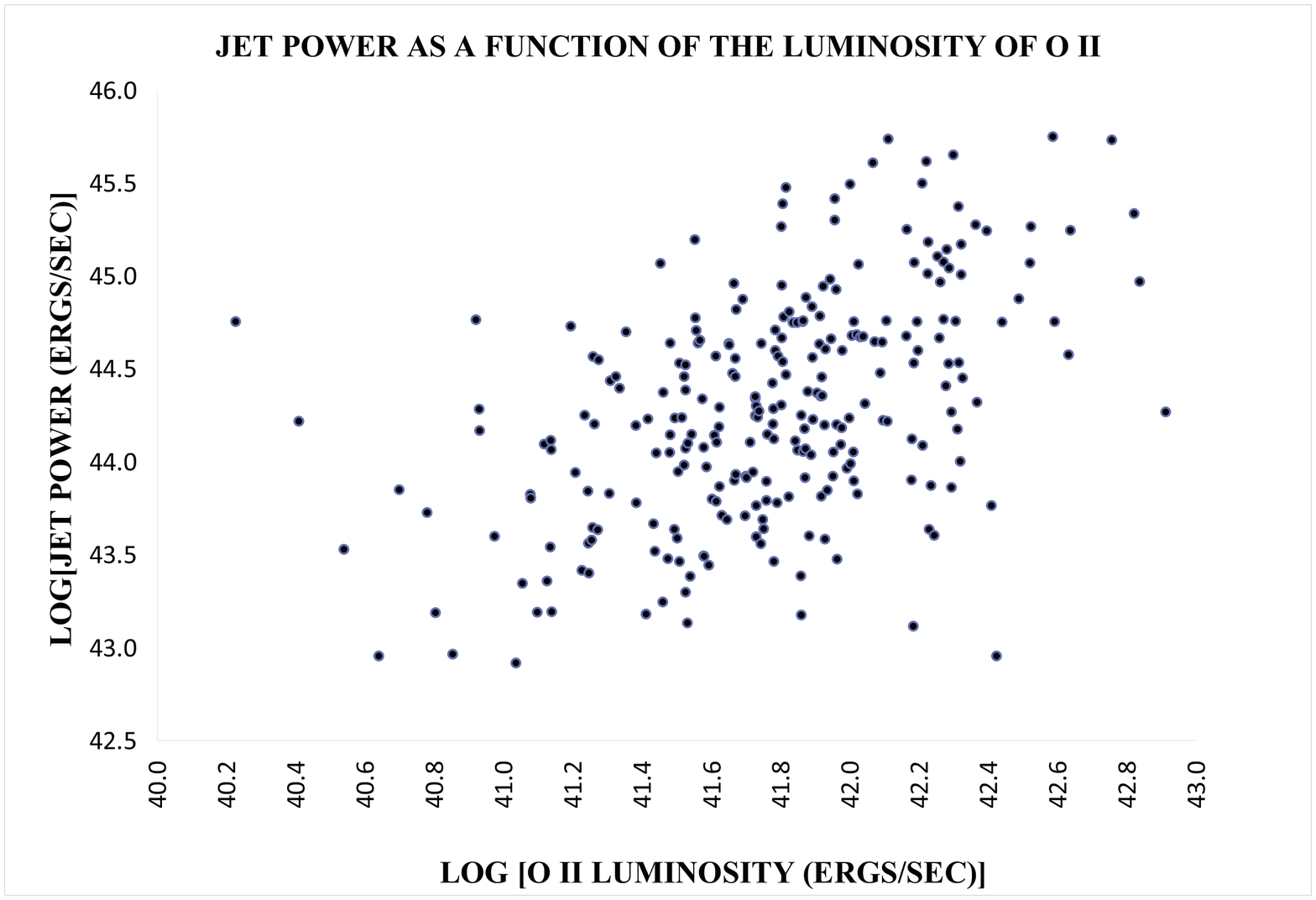}
\caption{On the left of the first row is the scatter plot of
$\log{\overline{Q}}$ versus $\log{L_{\mathrm{cont}}}$ and on the
right is the same for $\log{\lambda L_{\lambda}(3000\, \AA)}$ for
the 266 quasars with detected extended luminosity. Since the two
scatter plots look very similar, the effects of Doppler beaming must
be negligible within the sample as discussed in the text. Scatter
plots of the logarithm of the line strengths as a function of the
logarithm of $Q$ are in the next two rows. The narrow lines show a
modestly elevated level of correlation relative to the broad lines
in spite of some outliers with low line luminosity. The largest
scatter is with the continuum luminosity.}
\end{figure}
\section{Conclusion}
This Letter used a large SDSS DR7 sample and FIRST radio observations to address three fundamental questions
in the study of quasar radio jets. We repeat them here with answers indicated by the sample considered:
\begin{enumerate}
\item \textbf{How powerful is the typical radio jet?} From Figure 1, typically, the long term time averaged jet
power, $\overline{Q}$, is a factor of a few above the FR I/FR II
divide, $\overline{Q} \sim 1 \, -\, 5 \times 10^{44}\mathrm{ergs/sec}$.
\item \textbf{How frequently do quasars produce long term jets?} For $0.4 < z< 0.8$, 2.3\% of the optically selected quasars have FR II
level extended emission and $>0.4\%$ have FR I level extended emission indicative of long term jet production.
\item \textbf{How is the jet power related to the accretion flow thermal luminosity?} In the last section (see Figure 3), it was established
that the long term time average jet power, $\overline{Q}$, is not
strongly correlated with the accretion flow thermal luminosity. The
extended lobe emission used to compute $\overline{Q}$ was emitted
from the quasar central engine $\sim 10^{5}$ - $10^{7}$ years
earlier than the observed optical/UV continuum used to evaluate
$L_{bol}$ \citep{blu00}. Thus, the two properties are not
contemporaneously determined and there is no apriori reason to
expect them to correlate strongly, except possibly by the overall
scaling of the central black hole mass responsible for both. No
information was found in the sample that can be used to reliably
determine a connection between the instantaneous jet power and
$L_{bol}$.
\end{enumerate}

Another major finding of this work, that is closely related to point 3 above, is that
powerful FR II quasars ($\overline{Q} > 10^{45}\mathrm{ergs/sec}$) are extremely rare,
$\sim 0.3\%$ of all quasars. Thus, any theoretical model of the central engine must explain why
it is so difficult for a quasar environment to maintain a configuration conducive to strong jet
formation over $\sim 10^{5}$ - $10^{7}$ years.

\begin{acknowledgements}
We would like to thank the referee for many cogent and useful comments.
\end{acknowledgements}


\begin{thebibliography}{}
\bibitem[Antonucci(1993)]{ant93} Antonucci, R.J. 1993,
  Annu. Rev. Astron. Astrophys. \textbf{31} 473
\bibitem[Antonucci and Ulvestad(1985)]{ant85} Antonucci, R.J., Ulvestad, J. 1985,
  ApJ \textbf{294} 158
\bibitem[Blundell and Rawlings(2000)]{blu00} Blundell, K., Rawlings, S. 2000,
AJ \textbf{119} 1111
\bibitem[Blundell and Rawlings (2001)]{blu01} Blundell, K. and Rawlings, S. 2001, ApJL \textbf{562}
5
\bibitem[Buttiiglione et al.(2011)]{but11} Buttiglione, S. 2011, A \& A \textbf{525} 28
\bibitem[Buttiiglione et al.(2010)]{but10} Buttiglione, S. 2010, A \& A \textbf{509} 6
\bibitem[deVries et al.(2006)]{dev06} deVries, W., Becker, R., White, R. 2006, AJ \textbf{131} 666
\bibitem[Fernandes et al(2010)]{fer10} Fernandes, C. 2010, MNRAS in press arXiv:1010.0691v1
\bibitem[Gower and Hutchings(1984)]{gow84} Gower, A. and Hutchings, J. 1984, AJ
\textbf{89} 1658
\bibitem[Heckman et al(2004)]{hec04} Heckman, T. Kauffmann, G., Brinchmann, J.,Charlot, S, Tremonti, C.,
White, S.2004, ApJ \textbf{613} 109
\bibitem[Kaspi et al(2000)]{kas00} Kaspi, S. et al 2000, ApJ \textbf{533}
631
\bibitem[Kellermann \& Pauliny-Toth(1969)]{kel69} Kellermann,
K. I., \& Pauliny-Toth, I. I. K. 1969 ApJ, \textbf{155}, L71
\bibitem [Kellermann et al (1969)]{kel70}Kellermann, K. I., Pauliny-Toth, I. I. K., Williams, P. J. S. 1969 ApJ \textbf{157} 1
\bibitem[Koziel-Wierzbowska and Stasinska(2011)]{koz11}Koziel-Wierzbowska, D. and Stasinska, G.  2011, to appear in MNRAS, http://arxiv.org/abs/1101.3223v1
\bibitem[Malkan and Moore(1986)]{mal86}Malkan, M. and Moore, R. 1986, ApJ \textbf{300} 216
\bibitem[Murphy et al(1993)]{mur93} Murphy, D., Browne, I.,  Perley, R.1993, MNRAS \textbf{264} 298
\bibitem[Ogle et al(2006)]{ogl06} Ogle, P., Whysong, D. and Antonucci, R.2006 ApJ \textbf{647} 161
\bibitem[Punsly(2005)]{pun05}Punsly, B. 2005, ApJL \textbf{623} 9
\bibitem[Punsly(2008)]{pun08} Punsly, B. 2008, \emph{Black Hole
Gravitohydromagnetics}, second edition (Springer-Verlag, New York)
\bibitem[Punsly and Zhang (2010)]{pun10}Punsly, B., Zhang, S. 2010, ApJ \textbf{725} 1928
\bibitem[Punsly and Zhang (2011)]{pun11}Punsly, B., Zhang, S. 2011, MNRAS Letters \textbf{412} 123
\bibitem[Rawlings and Saunders(1991)]{raw91} Rawlings, S., Saunders,
  R. 1991, Nature \textbf{349}, 138
\bibitem[Simpson(1998)]{sim98}Simpson, C. 1998, MNRASL \textbf{297} 39
\bibitem[Smith et al(1988)]{smi88}Smith, P., Elston, R., Berriman, G., Allen, R.  1988, ApJL \textbf{326} 39
\bibitem[Sun and Malkan(1989)]{sun89}Sun, W.H., Malkan, M. 1989, ApJ \textbf{346} 68
\bibitem[Wang et al(2004)]{wan04}Wang, J.-M., Luo, B, Ho, L. 2004, ApJL \textbf{615} 9
\bibitem[Willott et al(1999)]{wil99}Willott, C., Rawlings, S.,
  Blundell, K., Lacy, M. 1999, MNRAS \textbf{309} 1017
\bibitem[Willott(2000)]{wil00}Willott, C., in \textbf{Proc. AGN in their Cosmic Environment", Eds. B. Rocca-Volmerange \& H. Sol, EDPS Conf. Series in Astron. \& Astrophysics} astro-ph/0007467
\bibitem[Zhang et al(2010)]{zha10}Zhang, S. et al 2010 ApJ \textbf{714} 367
\end{thebibliography}
\end{document}